\documentclass[aps,prd,twocolumn,showpacs,10pt,superscriptaddress,preprintnumbers,nofootinbib]{revtex4-1}
\pdfoutput=1

\usepackage[usenames,dvipsnames]{xcolor}
\usepackage{hyperref}

\usepackage{shuffle}
\usepackage{amsmath}
\usepackage{amsfonts}
\usepackage{mathtools}
\usepackage{microtype}
\usepackage{epsfig,amssymb,psfrag,epstopdf}
\usepackage{tikz}
\usepackage{bm}
\usepackage{array}
\usepackage{makecell}
\usepackage[utf8]{inputenc}
\usepackage{blkarray}
\usepackage{bigstrut}
\usepackage{nccmath}
\usepackage{enumitem}
\usepackage{braket}
\usepackage{dsfont}
\usepackage[english]{babel}
\usetikzlibrary{patterns,calc}
\usepackage{subfigure}
\usepackage{bookmark}
\usepackage{natbib}

\allowdisplaybreaks

\tikzset{
	dline/.style ={color = black, line width =2pt}
}
\tikzset{
	box/.style ={
		circle, 
		radius=10pt,
		inner sep=1pt, 
		draw=black }
}

\def\ket#1#2{\langle #1\,#2\rangle}
\def \be {\begin{equation}}
\def \ee {\end{equation}}

\begin{document}

\title{The KLT Relation from the Tree formula and Permutohedron}
\author{Qu Cao,}
\email{qucao@zju.edu.cn}
\author{Liang Zhang}
\email{liangzh@zju.edu.cn}
\affiliation{Zhejiang Institute of Modern Physics, Department of Physics, Zhejiang University, Hangzhou, 310027, China}

\begin{abstract}\noindent
In this paper, we generalize the Nguyen-Spradlin-Volovich-Wen (NSVW) \footnote{In their paper \cite{Nguyen:2009jk}, they admit that the formula they present is known from the older work by Bern, Dixon, Perelstein, and Rozowsky~\cite{Bern:1998sv}. So the tree formula should be called the BDPR formula. In our paper, we call the tree formula for convenience.} tree formula from the MHV sector to any helicity sector. We find a close connection between the Permutohedron and the KLT relation, and construct a non-trivial mapping between them, linking the amplitudes in the gauge and gravity theories. The gravity amplitude can also be  mapped from a determinant followed from the matrix-tree theorem. Besides, we use the binary tree graphs to manifest its Lie structure. In our tree formula, there is an evident Hopf algebra of the permutation group behind the gravity amplitudes. Using the tree formula, we can directly re-derive the soft/collinear limit of the amplitudes.
\end{abstract}

\maketitle

\section{Introduction}
\noindent The Kawai-Lewellen-Tye (KLT) relation~\cite{Kawai:1985xq}plays a pivotal role in scattering amplitudes for relating gravity amplitudes to gauge field amplitudes. In the field-theory limit, the string KLT relation reduces to the field KLT relation in a compact form \cite{Bern:1998sv}. The KLT relation is a kind of Double Copy \cite{Bern:2010ue}that originates from the Color-Kinematic duality~\cite{Bern:2008qj}. It uncovers the symmetry hidden in the Lagrangian of the two theories, i.e., the  gauge and gravity theory~\cite{Bern:1999ji}. The modern approach is expressed as the matrix form~\cite{Bjerrum-Bohr:2010diw,Bjerrum-Bohr:2010kyi}. The KLT matrix (or KLT kernel, Momentum Kernel) had been studied in string theory~\cite{Bjerrum-Bohr:2010pnr} until Cachazo, He, and Yuan (CHY)\cite{Cachazo:2013hca} found the inverse of the KLT matrix is the bi-adjoint $\phi^3$ amplitude, which has many geometric and combinatoric representations, such as the Associahedron~\cite{Arkani-Hamed:2017mur} and the intersection number~\cite{Mizera:2017cqs}. In recent years, we have seen original researches on the KLT relation \cite{Lin:2021pne,Chi:2021mio,Kalyanapuram:2020axt,Aoude:2019xuz,Barreiro:2019ncv,Gomis:2021ire,Li:2021yfk,Cho:2021nim,Guevara:2021tvr,Bonnefoy:2021qgu}, especially on the algebra structure \cite{Frost:2021qju,Frost:2020eoa}. However,  studies of the KLT relation  in  geometry and combinatorics are insufficient. This paper explores the KLT relation in these aspects by referring to the NSVW/BDPR tree formula~\cite{Nguyen:2009jk,Bern:1998sv} to discuss the tree structure of the KLT relation.  
The KLT relation is\footnote{We have omitted the $(-1)^n$ in the gravity amplitude.} 
\begin{equation}
	{\cal M}_n =\sum_{\alpha,\beta\in S_{n-3}} {\cal A}_n(1\alpha(n-1)n) {\cal S}[\alpha|\beta] {\cal A}_n(1\beta n(n-1))\,,
\end{equation}
where ${\cal M}_n$ is the $n$-point gravity amplitude, ${\cal A}_{n}$ is the color-ordered pure Yang-Mills amplitude\footnote{In the paper, we sometimes call  the gauge amplitude for convenience.}, $\alpha$ and $\beta$ are the permutations in the $S_{n-3}$ symmetry group, and ${\cal S}[\alpha|\beta]$ is the KLT matrix. When we choose $(n-3)!$ basis for the gauge amplitudes, the KLT matrix has a recursive
structure~\eqref{eq-KLT-recursive}, which can be used to derive the relation with the labelled trees (a brief proof in section $2.3.1.$ of the paper~\cite{Mafra:2020qst}). The recursive structure is
\be\label{eq-KLT-recursive}
{\cal S}[\alpha,j|\beta,j,\gamma]=2 p_{j}\cdot(p_{1}+p_{\beta}){\cal S}[\alpha|\beta,\gamma]\,,
\ee
where $s_{ij}=2p_{i}\cdot p_{j}$, and $p_{\beta}=\sum\limits_{i\in \beta}p_{i}$, and ${\cal S}[2|2]=s_{12}$.

\section{Start From The Tree Formula}
\noindent The  tree  formula uses the spanning trees to formulate the MHV gravity amplitudes~\cite{Nguyen:2009jk}.  In the $n$-point amplitudes, by fixing the point $n-1$,$n$, the remaining $n-2$ points  generate the spanning trees. Each edge has a weight like a propagator in the conditional Feynman rules. Gravity amplitudes can be derived by summing over the trees and multiplying an overall factor.
\begin{equation}
	\label{eq-NSVW}
	\begin{aligned}
		\mathcal{M}_{n}^{\mathrm{MHV}}=&\sum_{\text {trees}} \prod_{ edges~a b}{\frac{[ab]}{\langle a b\rangle}}\langle a(n-1)\rangle\langle b (n-1)\rangle\langle a n\rangle\langle b n\rangle \\&\times \frac{1}{\langle (n-1) n\rangle^{2}}\left(\prod_{a=1}^{n-2} \frac{1}{(\langle a (n-1)\rangle\langle a n\rangle)^{2}}\right)\,.
	\end{aligned}
\end{equation}

A natural question arises from the tree formula: how does one extend the gravity amplitudes tree formula beyond the MHV sector with the tree formula? To answer the question, the first step we need to do is that reformulate the  formula in a more general frame. At the tree level, the KLT relation is the best choice in this step. We should reformulate the tree formula to the KLT relation. The MHV pure Yang-Mills amplitude is the Parke-Taylor formula~\cite{Parke:1986gb} ${\cal A}_n^{\rm MHV}(\alpha(12\dots n))=\frac{1}{\prod_{i=1}^{n} \ket{\alpha(i)}{\alpha(i+1)}}$  \footnote{We use the convention that  ignores the common factor $\langle ij\rangle^{4}$ from the negative helicity particles in the MHV amplitudes.}. It is easy to use the Parke-Taylor formula to reformulate the  tree formula based on the KLT relation. 

For  $n=5$,  the gravity amplitude is the  sum of the below three tree graphs.
\begin{equation}
	\label{eq:fivepoint}
	\begin{aligned}
		{\hbox{\lower 5.pt\hbox{
					\begin{tikzpicture}
						\draw (14pt,0)--(28pt,0);
						\draw (42pt,0)--(56pt,0);
						\draw (7pt,0) circle (7pt);
						\node at (7pt,0) {1};
						\draw (35pt,0) circle (7pt);
						\node at (35pt,0) {2};
						\draw (63pt,0) circle (7pt);
						\node at (63pt,0) {3};
					\end{tikzpicture}
		}}}
		= s_{12}s_{23}{\cal A}_5^{\rm MHV}(12345){\cal A}_5^{\rm MHV}(12354)\,,
		\\
		{\hbox{\lower 5.pt\hbox{
					\begin{tikzpicture}
										\draw (14pt,0)--(28pt,0);
						\draw (42pt,0)--(56pt,0);
						\draw (7pt,0) circle (7pt);
						\node at (7pt,0) {1};
						\draw (35pt,0) circle (7pt);
						\node at (35pt,0) {3};
						\draw (63pt,0) circle (7pt);
						\node at (63pt,0) {2};
					\end{tikzpicture}
		}}}
		= s_{13}s_{23}{\cal A}_5^{\rm MHV}(13245){\cal A}_5^{\rm MHV}(13254)\,,
		\\
		{\hbox{\lower 5.pt\hbox{
					\begin{tikzpicture}
										\draw (14pt,0)--(28pt,0);
						\draw (42pt,0)--(56pt,0);
						\draw (7pt,0) circle (7pt);
						\node at (7pt,0) {2};
						\draw (35pt,0) circle (7pt);
						\node at (35pt,0) {1};
						\draw (63pt,0) circle (7pt);
						\node at (63pt,0) {3};
					\end{tikzpicture}
		}}}
		=s_{12}s_{13}{\cal A}_5^{\rm MHV}(21345){\cal A}_5^{\rm MHV}(21354)\,,
\end{aligned}
\end{equation}
which is consistent with the KLT relation for $n=5$, after we replace ${\cal A}_5^{\rm MHV}(21345) {\cal A}_5^{\rm MHV}(21354)$ with the basis ${\cal A}_5^{\rm MHV}(12345),{\cal A}_5^{\rm MHV}(12354),{\cal A}_5^{\rm MHV}(13245),{\cal A}_5^{\rm MHV}(13254)$. The replacement is
\begin{equation}
	\label{eq-MHV-5pt-id}
	\begin{aligned}
		&{\cal A}_5^{\rm MHV}(21345){\cal A}_5^{\rm MHV}(21354)\\
		&=\frac{\ket{1}{4}}{\ket{1}{2}\ket{1}{3}\ket{2}{4}\ket{3}{4}\ket{4}{5}\ket{5}{1}}\frac{\ket{1}{5}}{\ket{1}{2}\ket{1}{3}\ket{2}{5}\ket{3}{5}\ket{5}{4}\ket{4}{1}}\\
		&=({\cal A}_5^{\rm MHV}(1(2\shuffle3)45))({\cal A}_5^{\rm MHV}(1(2\shuffle3)54))\\
		&=({\cal A}_5^{\rm MHV}(12345)+{\cal A}_5^{\rm MHV}(13245))\\
		&\quad\times({\cal A}_5^{\rm MHV}(12354)+{\cal A}_5^{\rm MHV}(13254))\,,
	\end{aligned}
\end{equation}
 where $\shuffle$ denotes the shuffle, $(a\shuffle b)=\{ab\}+\{ba\}$. 
 
  From the above example, it can be learned that if we want to get a formula based on the KLT relation, we should use the ${\cal A}(1 \alpha (n-1) n)$ and ${\cal A}(1 \beta n (n-1))$ to form the tree formula.  From here, we make a convention of the tree graph that it only represents the KLT matrix, not the gravity amplitude. Each edge denotes the Mandelstam variables $s_{ij}$. We choose point $1$ as the root to form the rooted labelled tree in the KLT matrix.



Take the five-point amplitude as an example. 
\begin{align}
	\mathcal{M}_5
	&=\begin{pmatrix}
	    A_1,A_2
	\end{pmatrix}
	\begin{pmatrix}
	    s_{12}(s_{13}+s_{23}) & s_{12}s_{13}\\
	    s_{12}s_{13} & s_{13}(s_{12}+s_{23})
	\end{pmatrix}
	\begin{pmatrix}
	    \tilde{A}_1\\
	    \tilde{A}_2
	\end{pmatrix}
\end{align}
Then we denote the $s_{ij}$ as the tree graph :\put(4,-5){\tikz{	\draw (10pt,0)--(20pt,0);
					\draw (5pt,0) circle (5pt);
					\node at (5pt,0) {i};
					\draw (25pt,0) circle (5pt);
					\node at (25pt,0) {j};}}
\begin{equation}
    \begin{aligned}\label{eq-M5-tree}
	{\cal M} _5&=\begin{pmatrix}
		A_1,A_2 
	\end{pmatrix}
	\begin{pmatrix}
		\begin{tikzpicture}[sibling distance =25pt,scale=0.5]
			\node[box] {1}
			child {node[box] {2}}
			child {node[box] {3}};
		\end{tikzpicture} \put(-2,15){+}  \quad\begin{tikzpicture}[sibling distance =20pt,scale=0.4]
			\node[box] {1}
			child {node[box] {2}}{
				child {node[box] {3}}};
		\end{tikzpicture} &\quad
		\begin{tikzpicture}[sibling distance =25pt,scale=0.5]
			\node[box] {1}
			child {node[box] {2}}
			child {node[box] {3}};
		\end{tikzpicture}\\
		\begin{tikzpicture}[sibling distance =25pt,scale=0.5]
			\node[box] {1}
			child {node[box] {2}}
			child {node[box] {3}};
		\end{tikzpicture} & \begin{tikzpicture}[sibling distance =25pt,scale=0.5]
			\node[box] {1}
			child {node[box] {2}}
			child {node[box] {3}};
		\end{tikzpicture} \put(-2,20) {+}  \quad\begin{tikzpicture}[sibling distance =20pt,scale=0.4]
			\node[box] {1}
			child {node[box] {3}}{
				child {node[box] {2}}};
		\end{tikzpicture} 
	\end{pmatrix}
	\begin{pmatrix}
		\tilde{A_1}\\
		\tilde{A_2}
	\end{pmatrix}\\
	&= {\hbox{\lower 20.pt\hbox{\begin{tikzpicture}[sibling distance =25pt,scale=0.6]
					\node[box] {1}
					child {node[box] {2}}
					child {node[box] {3}};
	\end{tikzpicture}}}}{\cal A}(1 (2\shuffle 3) 4 5){\cal A}(1 (2\shuffle 3) 5 4) \\&\quad+{\hbox{\lower 20.pt\hbox{\begin{tikzpicture}[sibling distance =20pt,scale=0.5]
					\node[box] {1}
					child {node[box] {2}}{
						child {node[box] {3}}};
	\end{tikzpicture}}}}{\cal A}(1 2 3 4 5){\cal A}(1 2 3 5 4)\nonumber
	+ {\hbox{\lower 20.pt\hbox{\begin{tikzpicture}[sibling distance =20pt,scale=0.5]
					\node[box] {1}
					child {node[box] {3}}{
						child {node[box] {2}}};
	\end{tikzpicture}}}}{\cal A}(1 3 2 4 5){\cal A}(1 3 2 5 4)\,,
\end{aligned}
\end{equation}
where $A_{1}={\cal A}(12345)$, $A_2={\cal A}(13245)$, $\tilde{A_1}={\cal A}(12354)$, $\tilde{A_2}={\cal A}(13254)$. This formula becomes the  tree formula~\eqref{eq:fivepoint} in the MHV sector.

In conclusion, the gravity amplitudes are expanded by the tree graphs. The $(n-3)!\times (n-3)!$ KLT matrix has the tree structure, which can reduce the number of independent elements from $\frac{((n-3)!+1)(n-3)!}{2}$ to $(n-2)^{(n-4)}$ for  Cayley's formula~\cite{cayley1889theorem}. 
\begin{equation}
	\label{eq-KLT-tree-formula}
	{\cal M}_n = \sum\limits_{g}\sum_{\alpha,\beta}  \prod\limits_{(ij)\in E(g)}s_{ij} {\cal A}_n(1\alpha(n-1)n){\cal A}_n(1\beta n(n-1))\,,
\end{equation}
where the permutation sets $\alpha$, $\beta$ belong to the $S_{n-3}$, $g$ is the compatible graph. $(ij)$ is the edge connected with vertex $i$ and $j$, $E(g)$ is the set of the edges of the $g$ graph. We will discuss in details and give the proof in ~\ref{section-proof}.

\section{The Determinant Of Matrix}
\noindent It is well known that the tree graphs can be related to a determinant of the matrix for the matrix-tree theorem~\cite{Feng:2012sy}. In the MHV sector, the  tree formula can be derived from the Hodges formula~\cite{Hodges:2012ym}. Since we get the tree formula based on the KLT relations, a natural question arises about how to derive a determinant for the gravity amplitude, i.e., how to generalize the Hodges formula. The idea is similar to the tree formula, and we do not expect the whole gravity amplitudes can be easily implemented as a determinant.  The formula~\eqref{eq-KLT-tree-formula} becomes a sum of the trees with edge $s_{ij}$ when the gauge amplitudes equal one, i.e., ${\cal A}_n=1$. This is a hint for us  to generalize the Hodges formula using the matrix-tree theorem.

The first step in seeking the determinant representation is to build a weighted Laplacian matrix~\cite{Feng:2012sy}. Following the convention above is an obvious choice for us to construct the matrix. The off-diagonal entries are the product of $s_{ij}$ and a list $\{ij\}$ connected with the point $i$ and $j$.
\be
W_{ij}=s_{ij}\{ij\}\equiv  \psi_{i j} ,\qquad W_{ii}=-\sum_{i\neq j} \psi_{i j}.
\ee

We use the matrix-tree theorem to expand the weighted Laplacian matrix.
\be
|W(G)|_{i}^{i}=\sum_{T \in \mathcal{T}(G)}\left(\prod_{e=\left(v_{i} v_{j}\right) \in E(T)} s_{ij}\{ij\}\right),
\ee
where the connected, simple graph $G$ with vertices $V=\left\{v_{1}, \ldots v_{n-2}\right\}$, the sum is over all spanning trees $T \in \mathcal{T}(G)$, the product is over all edges of $e \in T $, and $|W(G)|_{i}^{i}$ denotes the determinant of the matrix without the i-th row and i-th column. The choice of the $i$ is arbitrary.

The determinant is not the gravity amplitude yet, and we need one more step for defining a map to make the lists become the gauge amplitudes in the $(n-3)!$ basis. 
\be\label{K-map}
{\cal K}: {\alpha} \rightarrow {\cal A}(\alpha (n-1) n){\cal A}(\alpha n (n-1))\,,
\ee
where $\alpha$ is the list from $1$ to $n-2$, generated by the $\{ij\}$.

Therefore, the gravity amplitudes are mapped from a determinant of the matrix,
\be{}
{\cal M}_{n}={\cal K}(|W(G)|_{i}^{i})\,.
\ee{}

From this formula, each component of gravity amplitude is a product of tree graph and the gauge amplitude. When the degree of vertex $j$ in the graph, $j\neq1$, is greater than two, the list in the gauge amplitude will be a shuffle. For the root $1$, a degree greater than one suffices. For example, the degree of vertex 2 in the Appendix~\ref{app}  (A2) is three, then the gauge amplitudes followed it are ${\cal A}(1 2(3\shuffle 4) 5 6){\cal A}(1 2(3\shuffle 4) 6 5)$ for the map ${\cal K}$~\eqref{K-map}. In this case, the list $\alpha$ is $\{12(3\shuffle4)\}$, comes from $\{12\}$, $\{23\}$, $\{24\}$.

\section{The KLT Permutohedron}
\noindent The Permutohedron \cite{Williams:2015twa,Early:2017lku,Early:2018zuw} is the graphic representation of the symmetry group $S_{n}$, consisting of the $n!$ vertices of the permutation of the order $n$. It is denoted by ${\cal P}^{n}$ with the dimension $(n-1)$. The gravity amplitude is also $S_{n}$-invariant , so there must exist some direct connections between the amplitude and the Permutohedron. The amplitude can be reduced to  $S_{n-3}$ with the KLT matrix. In general, the $n$-point amplitudes correspond to ${\cal P}^{n}$ restricted to the $n-3$ dimension. 

For example, the dimension of the ${\cal P}^{4}$ is $3$, and the amplitude can be mapped from the codimension $2$ facet of the Permutohedron ${\cal P}^{4}|_{1}$ , which is an edge connecting permutation $\{1234\}$ and $\{1243\}$. The permutations map to the gauge amplitudes ${\cal A}(1234)$ and ${\cal A}(1243)$. The edge, which denotes the transposition, maps to the Mandelstam variable $s_{12}$. Then, the edge becomes the ${{\cal M}_{4}}=s_{12}{\cal A}(1234){\cal A}(1243)$. The map is 
\be\label{eq-permutohedron}
\Phi: {\cal P}^{n} \rightarrow {\cal M}_{n}, \quad \mathcal{M}_n=\Phi({\cal P}^{n})|_{n-3}\,,
\ee
which means the amplitude can be retained by the map from the codimension $2$ boundaries of the Permutohedron. 

The construction of the map is not trivial for some shuffle structures of the amplitudes, which we learned from the above section. Each vertex represents one ${\cal A}(\alpha)$. The mapping rule is shown in the table \ref{tab:my_label},
\begin{table}[ht]
    \centering
    \caption{Map Rule}
    \label{tab:my_label}
\begin{tabular}{|c|c|c|c|m{1.5cm}<{\centering}|}
\hline 
Number&${\cal P}^n|_{n-3}$&$S_{i}|_{\tau}$& (p,q)& ${\cal T}_{(p,q)}$\\
\hline	
\makecell*[c]{1}&\makecell*[c]{\tikz{\fill(0,0)circle[radius=2pt];}}&$\makecell*[c]{S_1}$&\makecell*[c]{(0,1)}&\makecell*[c]{\tikz{\draw[dashed] (0,0)--(0,-.6);
\fill (0,-.7)circle[radius=2pt];}}\\
\hline
\makecell*[c]{2}&\makecell*[c]{\tikz{\filldraw (0,0)circle[radius=2pt]--(1,0)circle[radius=2pt];}}&\makecell*[c]{$S_2$}&\makecell*[c]{(1,1)}&\makecell*[c]{\tikz{
    \draw[dashed] (0,-0.4)--(0,-1);
	\draw (0,-1)++(-60:.6)--(0,-1)--++(-120:.6);
	\fill (0,-1)++(-60:.6)circle[radius=2pt];
	\fill (0,-1)++(-120:.6)circle[radius=2pt];}}\\
\hline
\makecell*[c]{3}&\makecell*[c]{\tikz\filldraw (0,0)circle[radius=2pt]--(1,0)circle[radius=2pt]--(-60:1)circle[radius=2pt]--(0,0);}&\makecell*[c]{$S_3|_\tau$}&\makecell*[c]{(1,2)}&\makecell*[c]{\tikz{
    \draw[dashed] (0,-0.4)--(0,-1);
	\draw (0,-1)++(-30:.6)--(0,-1)--++(-150:.6);
	\fill (0,-1)++(-150:.6)circle[radius=2pt];
	\filldraw (0,-1)++(-30:.6)circle[radius=2pt]--++(0,-0.6)circle[radius=2pt];}}\\
\hline
\makecell*[c]{\vdots}&&&&\\
\hline
\makecell*[c]{$(n-3)!$}&$\makecell*[c]{{\cal P}^n|_{n-3}}$&\makecell*[c]{$S_{n-3}$}&\makecell[b]{$\underbrace{(1,\cdots,1)}_{n-3}$}&\makecell*[c]{\tikz{
	\filldraw (-150:.6)circle[radius=2pt]--(0,0)circle[radius=2pt]--(-30:.6)circle[radius=2pt];
	\filldraw (-60:.6)circle[radius=2pt]--(0,0)circle[radius=2pt]--(-120:.6)circle[radius=2pt];
	\draw[dotted,thick](-75:.6)arc(-75:-105:.6)}}\\
\hline	
\end{tabular}
\end{table}

The shuffle form $(p,q)$ is a shuffle between the length $p$ list and the length $q$ list, which has the number  ${p+q \choose p}$, the shuffle trees denote ${\cal T}_{(p,q)}$, and the $S_{i}|{\tau}$ denotes the permutation group restricted to the ordered list $\tau$. Each ${\cal A}(1 \alpha (n-1) n)$ and ${\cal A}(1 \alpha n (n-1))$ are mapped from codimension $3$ boundaries, and the KLT matrix connects them as a bridge to form the ${\cal P}^n|_{n-3}$. We call ${\cal P}^n|_{n-3}$ as the KLT Permutohedron.

$\mathcal{M}_5$ can be mapped from the $\mathcal{P}^{5}|_{2}$,  a rectangle, of which each vertex represents a gauge amplitude. The edge on the top/bottom represents the path graph. The whole rectangle represents the star graph ${\cal T}_{(1,1)}$. 

\begin{tikzpicture}
	\node at (-1,0.6) {$\mathcal{M}_5:$};
	\draw[pattern=north east lines,pattern color=teal] (0,0) rectangle (3,1.5);
	
	\path(1.5,0.2) node[above]{{\hbox{\lower 10.pt\hbox{\begin{tikzpicture}[sibling distance =35pt,scale=0.5]
						\node[box] {1}
						child {node[box] {2}}
						child {node[box] {3}};
	\end{tikzpicture}}}}};
	\draw[violet,thick] (0,0) --(3,0);
	\draw[red,thick] (0,1.5) --(3,1.5);
	\fill (0,1.5)circle[radius=2pt]node [above]{$A(12345)$};
	\fill(0,0)circle[radius=2pt] node[below]{$A(13245)$};
	\fill(3,1.5)circle[radius=2pt] node[above]{$A(12354)$};
	\fill(3,0)circle[radius=2pt] node[below]{$A(13254)$};
\end{tikzpicture}
\begin{tikzpicture}
	\draw[red] (0,0)--(1,0);
	\path(0.5,0)node[below]{\begin{tikzpicture}[sibling distance =35pt,scale=0.3]
			\node[box] {1}
			child {node[box] {2}}{
				child {node[box] {3}}};
	\end{tikzpicture}};
\end{tikzpicture}
\hspace{0.5cm}
\begin{tikzpicture}
	\draw[violet] (0,0)--(1,0);
	\path(0.5,0)node[below]{\begin{tikzpicture}[sibling distance =35pt,scale=0.3]
			\node[box] {1}
			child {node[box] {3}}{
				child {node[box] {2}}};
	\end{tikzpicture}};
\end{tikzpicture}
Once we sum all the contributions from the map of the Permutohedron  $\mathcal{P}^{5}|_{2}$, we get the five-point gravity amplitude in~\eqref{eq:fivepoint}.

${\cal M}_{6}$ is mapped from the ${\cal P}^{6}|_{3}$, which is restricted to the dimension $3$ part between permutation $\{1\alpha56\}$ and $\{1\alpha65\}$. All tree graphs come from the dimension $1$, $2$, and $3$ of the Permutohedron, made up of the KLT matrix in six points. The Permutohedron $\mathcal{P}^{6}|_{3}$ is a hexagonal prism. Each vertex represents a gauge amplitude.    \\
\\
\begin{minipage}{0.4\linewidth}
$\mathcal{M}_6$:
{\hbox{\lower 50.pt\hbox{\begin{tikzpicture}
							    \draw (0:1)node(R1-0){}--(300:0.5)node(R1-315){}--(210:0.87)node(R1-225){}--(180:1)node(R1-180){};
			     \draw[densely dashed,thick] (0:1)--(30:0.87)node(R1-45){}--(120:0.5)node(R1-135){}--(180:1);
				\begin{scope}[yshift=2.cm]
			    \draw (0:1)node(R2-0){}--(300:0.5)node(R2-315){}--(210:0.87)node(R2-225){}--(180:1)node(R2-180){};
			     \draw (0:1)--(30:0.87)node(R2-45){}--(120:0.5)node(R2-135){}--(180:1);
		     \end{scope}
				
				\fill[orange!20] (R1-0.center)--(R1-45.center)--(R2-45.center)--(R2-0.center)--cycle;
				\fill[yellow!20] (R1-135.center)--(R1-180.center)--(R2-180.center)--(R2-135.center)--cycle;
				\fill[red!20] (R1-225.center)rectangle(R2-315.center);
				\draw[Black] (R2-180.center)--(R2-225.center);
				\draw[Black] (R2-315.center)--(R2-0.center);
				\draw[RedViolet,thick] (R1-0.center)--(R2-0.center);
				\draw[blue,thick] (R1-180.center)--(R2-180.center);
				\draw[LimeGreen,thick] (R1-225.center)--(R2-225.center);
				\draw[Green,thick] (R1-315.center)--(R2-315.center);
				\draw[Magenta,dashed,thick] (R1-45.center)--(R2-45.center);
				\draw[SkyBlue,dashed,thick] (R1-135.center)--(R2-135.center);
				\fill[blue] (R1-0.center)node[anchor=west]{$A$}circle[radius=2pt];
				\fill[violet] (R1-45.center)node[anchor=south west]{$B$}circle[radius=2pt];
				\fill[Emerald] (R1-135.center)node[anchor=south]{$C$}circle[radius=2pt];
				\fill[WildStrawberry] (R1-180.center)node[anchor=east]{$D$}circle[radius=2pt];
				\fill[YellowOrange] (R1-225.center)node[anchor=north]{$E$}circle[radius=2pt];
				\fill[RawSienna] (R1-315.center)node[anchor=west]{$F$}circle[radius=2pt];
				\fill[blue] (R2-0.center)node[anchor=west]{$A^\prime$}circle[radius=2pt];
				\fill[violet] (R2-45.center)node[anchor=south west]{$B^\prime$}circle[radius=2pt];
				\fill[Emerald] (R2-135.center)node[anchor=south]{$C^\prime$}circle[radius=2pt];
				\fill[WildStrawberry] (R2-180.center)node[anchor=east]{$D^\prime$}circle[radius=2pt];
				\fill[YellowOrange] (R2-225.center)node[anchor=north]{$E^\prime$}circle[radius=2pt];
				\fill[RawSienna] (R2-315.center)node[anchor=west]{$F^\prime$}circle[radius=2pt];
				\draw[dashed] (30:0.87)--(120:0.5);
				\draw[dashed] (180:1)--(120:0.5);
				\draw (R2-225.center)--(R2-315.center);
				\draw (R2-0.center)--(R2-45.center);
				\draw (R2-135.center)--(R2-180.center);
\end{tikzpicture}}}}
\end{minipage}\hfill
\begin{minipage}{0.6\linewidth}
A : ${\cal A}$(142356)\quad$A^\prime : {\cal A}$(142365)\\
B : ${\cal A}$(143256)\quad$B^\prime : {\cal A}$(143265)\\
C : ${\cal A}$(134256)\quad$C^\prime : {\cal A}$(134265)\\
D : ${\cal A}$(132456)\quad$D^\prime : {\cal A}$(132465)\\
E : ${\cal A}$(123456)\quad$E^\prime : {\cal A}$(123456)\\
F : ${\cal A}$(124356)\quad$F^\prime : {\cal A}$(124356)		
\end{minipage}\\

The six-point gravity amplitude can be derived from the sum of the map of the Permutohedron ${\cal P}^6|_{3}$. See the details in  Appendix~\ref{app}.\\

\section{The Lie structure and binary tree} 
\label{section-bt}
\noindent The KLT matrix originates from the string theory when calculating the closed string amplitudes. Each $s_{ij}$ comes from the discontinuity of the Koba Nielsen factors \cite{Bjerrum-Bohr:2010pnr,Sondergaard:2011iv}.When the string tension gets to infinite, i.e., $\alpha^{\prime}\to0$, the KLT matrix forms in the field limit.
\be
\frac{e^{i\pi\alpha^{\prime}p_{i}\cdot p_{j}}-e^{-i\pi\alpha^{\prime}p_{i}\cdot p_{j}}}{2i}=sin(\pi\alpha^{\prime}p_{i}\cdot p_{j}) \rightarrow p_{i}\cdot p_{j}\footnote{Here in the field-theory limit, we have omitted $\alpha^{\prime}$ in the expression.}\,,
\ee

The discontinuity has the Lie structure, similar to the study of the Lie Polynomials in \cite{Frost:2019fjn,Frost:2020eoa}. The KLT matrix  diagonal can be expressed as the Lie Polynomials or the binary tree graphs under the map, similar to the study of the bi-adjoint $\phi^{3}$ amplitudes in \cite{Mafra:2016ltu}.
\be\label{eq-binary}
{\cal S}[\alpha|\alpha]={\cal L}([[[[1,\alpha_{2}],\alpha_{3}],\dots],\alpha_{n-2}])\,,
\ee
where the ${\cal L}$ is a mapping, ${\cal L}: Lie\quad Polynomials \rightarrow Kinematic\quad Space$. It has a recursive definition that ${\cal L}([\alpha,j])={\cal L}(\alpha)\Phi(\alpha,j)$,and $\alpha$ is the Lie Polynomials. $i$ and $j$ are letters, $\Phi(i,j)=2p_{i}\cdot p_{j}$ in the field-theory limit, and $\Phi(i,j)=sin(\pi\alpha^{\prime}p_{i}\cdot p_{j})$ in string theory. As follows, we use the binary tree graphs to express the Lie structure manifestly \cite{garsia1990combinatorics}.\\
For example, ${\cal S}[234|234]$=\put(0,-35){\tikz{
		\draw (135:1)node at (135:1.1){1}--(0,0)--(0:1)--(0:2)--++(45:1)node at ++(45:.2){5,6};
		\draw (0,0)--(-135:1)node at (-135:1.2){2}(0:1)--++(-90:1)node at ++(-90:.2){3}(0:2)--++(-45:1)node at ++(-45:.2){4};
		\fill[blue] (0,0)circle[radius=2pt];
		\fill[violet] (0:1)circle[radius=2pt];
		\fill[green](0:2)circle[radius=2pt];}}\\
where each line has the momentum and obeys the conservation of momentum. Each vertex has the factor $2p_{1}\cdot p_{2}$, $2(p_{1}+p_{2})\cdot p_{3}$ and $2(p_{1}+p_{2}+p_{3})\cdot p_{4}$.
\begin{equation}
	{\cal S}[234|234]=s_{12}(s_{13}+s_{23})(s_{14}+s_{24}+s_{34})\,.
\end{equation}

${\cal S}[\alpha|\alpha]$ can be represented by these binary tree graphs or a toy model of the on-shell Feynman graph with the vertex interaction but no propagators.\\
\begin{tikzpicture}
	\draw (135:1)node at (135:1.1){1}--(0,0)--(0:1)--(0:2)(0:3.5)--(0:4.5);
	\draw (0,0)--(-135:1)node at (-135:1.2){$\alpha(2)$}(0:1)--++(-90:1)node at ++(-90:.2){$\alpha(3)$}(0:2)--++(-90:1)node at ++(-90:.2){$\alpha(4)$}
	(0:3.5)--++(-90:1)node at ++(-90:.2){$\alpha(n-1)$};
	\draw[thick,dotted] (0:2)--(0:3.5);
	\draw (0:4.5)--++(-45:1)node at ++(-45:.2){$\alpha(n-2)$};	
	\draw (0:4.5)--++(45:1)node at ++(45:.2){$n-1,n$};
	\fill[red] (0,0)circle[radius=2pt];
	\fill[red] (0:1)circle[radius=2pt];
	\fill[red] (0:2)circle[radius=2pt];
	\fill[red] (0:4.5)circle[radius=2pt];
\end{tikzpicture}\\
In the ground of the binary tree graphs, $S[\alpha|\beta]$ can be treated as the intersection of two graphs, leading to the equation \eqref{eq-s-tree}.\\
\begin{align}
	\begin{Bmatrix}
		\begin{tikzpicture}[scale=0.8]
			\draw (135:1)node at (135:1.1){1}--(0,0);
			\draw (0,0)--(-135:1)node at (-135:1.3){$\alpha(2)$}
			;
			\draw[dotted,thick] (0:0)--(0:1);
			\draw (0.2,0)--(0.2,-1);
			\draw (0.8,0)--(0.8,-1);
			\draw[dotted,thick] (0.3,-1)--(0.7,-1);
			\draw (0:1)--++(-45:1)node at ++(-45:.3){$\alpha(n-2)$};
			\draw (0:1)--++(45:1)node at ++(45:.3){$n-1,n$};		
		\end{tikzpicture} & , \begin{tikzpicture}[scale=0.8]
			\draw (135:1)node at (135:1.1){1}--(0,0);
			\draw (0,0)--(-135:1)node at (-135:1.3){$\beta(2)$}
			;
			\draw[dotted,thick] (0,0)--(1,0);
			\draw (0.2,0)--(0.2,-1);
			\draw (0.8,0)--(0.8,-1);
			\draw[dotted,thick] (0.3,-1)--(0.7,-1);
			\draw (0:1)--++(-45:1)node at ++(-45:.3){$\beta(n-2)$};
			\draw (0:1)--++(45:1)node at ++(45:.3){$n-1,n$};		
		\end{tikzpicture}
	\end{Bmatrix}
\end{align}

\section{The proof of tree formula}\label{section-proof}
\noindent The elements of the inverse of the KLT matrix are the bi-adjiont $\phi^{3}$ amplitudes \cite{Cachazo:2013iea,Mizera:2016jhj},\\
\be{\label{eq-fi3}}
m_{\phi^3}(\alpha|\beta)=(-1)^{flip(\alpha|\beta)}\sum\limits_{g\in T(\alpha)\cap T(\beta)}\prod\limits_{I\in p(g)}\frac{1}{s_{I}}\,,
\ee
where $T(\alpha)$ denotes the binary tree graphs \cite{Gao:2017dek} compatible with $\alpha$, $p(g)$ is the set of the propagators of $g$ graph, $ s_I = (\sum\limits_{i \in I} p_i)^2$ . 

By analogy, we propose the formula for the KLT matrix, which is proved from the recursive structure or the binary tree representation in \ref{section-bt}.
\be{\label{eq-s-tree}}
S[\alpha|\beta]=\sum\limits_{g\in F(\alpha)\cap F(\beta)}\prod\limits_{(ij)\in E(g)}s_{ij}\,,
\ee
where $F(\alpha)$ denotes the set of all tree graphs compatible with $\alpha$. The compatible tree graphs mean that the rooted labelled trees can become the ordered lists with some shuffle operation. $E(g)$ is the set of the edges of the $g$ graph. 

Here is an example for the ${\cal S}[\alpha|\beta]$,
\begin{align}
	&F(23)=\begin{Bmatrix}
		\begin{tikzpicture}[sibling distance =30pt,scale=0.45]
			\node[box] {1}
			child {node[box] {2}}{
				child {node[box] {3}}};\end{tikzpicture} & , & \begin{tikzpicture}[sibling distance =25pt,scale=0.6]
			\node[box] {1}
			child {node[box] {2}}
			child {node[box] {3}};
		\end{tikzpicture} \quad
	\end{Bmatrix}\,,\quad 
		F(32)=\begin{Bmatrix}
		\begin{tikzpicture}[sibling distance =30pt,scale=0.45]
			\node[box] {1}
			child {node[box] {3}}{
				child {node[box] {2}}};
		\end{tikzpicture} & , & \begin{tikzpicture}[sibling distance =25pt,scale=0.6]
			\node[box] {1}
			child {node[box] {3}}
			child {node[box] {2}};
		\end{tikzpicture}
	\end{Bmatrix}\,. \\
		&g=F(23)\cap F(32)=\put(0,-10){\begin{tikzpicture}[sibling distance =40pt,scale=0.4]
		\node[box] {1}
		child {node[box] {3}}
		child {node[box] {2}};
	\end{tikzpicture}}\qquad \quad ,\quad{\cal S}[23|32]=s_{12}s_{13}.
\end{align}

Using the \eqref{eq-s-tree}, we can easily prove the tree formula for the KLT relation~\eqref{eq-KLT-tree-formula}.
Here $g$ belongs to the $ F(\alpha)\cap F(\beta)$.
\begin{equation}
\begin{aligned}
	\label{eq-KLT-tree-formula2}
	{\cal M}_n =&\sum_{\alpha,\beta\in S_{n-3}} {\cal A}_n(1\alpha(n-1)n) {\cal S}[\alpha|\beta] {\cal A}_n(1\beta n(n-1))\\ 
	=&\sum\limits_{g}\sum_{\alpha,\beta}  \prod\limits_{(ij)\in E(g)}s_{ij} {\cal A}_n(1\alpha(n-1)n){\cal A}_n(1\beta n(n-1))\,.
\end{aligned}
\end{equation}

The traditional KLT formula is that sum over the permutation sets $\alpha$, $\beta$, and now we change to sum over the tree graphs with the edges $s_{ij}$ and corresponding gauge amplitudes ${\cal A}_n(1\alpha(n-1)n){\cal A}_n(1\beta n(n-1))$, which will appear a shuffle operation as same as the we have seen in the five points~\eqref{eq-MHV-5pt-id}. The origin of these shuffle structures can be viewed as a hidden Hopf algebra discussed in the next section.

\section{The Hopf algebra}
\noindent The Hopf algebra has been studied in the scattering amplitudes \cite{Duhr:2012fh}. The tree formula and the shuffle structure in the KLT relation imply  a Hopf algebra exists. The MPR Hopf algebra is a Hopf algebra of the permutation group \cite{malvenuto1995duality}, so the permutation group can be mapped to the color-ordered amplitudes while keeping the Hopf structure. We define the $\mathbb{Q}$-vector space $H$ as the infinite sum of $H_{n}$ space, 
\begin{equation}
H=\bigoplus_{n=0}^{\infty} H_{n}=H_{0} \oplus H_{>0}\,, \quad H_{0}=\mathbb{Q}\,,
\end{equation}
where the $S_{n}$ permutation groups belong to the $H_{n}$ space. The coproduct $\Delta$ of the Hopf algebra is the shuffle $\shuffle$, which keeps the grading of the Hopf algebra,
\begin{equation}
\Delta\left(H_{n}\right) \subseteq \bigoplus_{p+q=n} H_{p} \otimes H_{q}\,,
\end{equation}
and we can define the iterated coproduct,
\be
\Delta_{i_{1}, \ldots, i_{k}}: H \rightarrow H_{i_{1}} \otimes \cdots \otimes H_{i_{k}}\,.
\ee
and define a pullback reflection,
\be
{\cal C}_{n}:  H_{i_{1}} \otimes \cdots \otimes H_{i_{k}} \rightarrow H_{n}
\ee
where ${\cal C}_{n}(\alpha_{i_{1}}\shuffle\cdots\shuffle\alpha_{i_{k}}) =(\alpha_{n-i})(\alpha_{i_{1}}\shuffle\cdots\shuffle\alpha_{i_{k}})$, $\alpha_{i_{k}}$ is the permutation list of the length $i_{k}$, and  $i=i_{1}+\ldots i_{k}$.

We map the permutation group to the gauge amplitudes. The map ${\cal Z}:H\rightarrow \tilde{H}$,  and $\tilde{H}$ is the vector space of the gauge amplitudes. Then the amplitude can be generated by the iterated coproduct and pullback  of the $\tilde{H}$,
\be
\begin{aligned}
 &{\cal A}(1\alpha)\subset {\cal C}_{n-2}^{1}\Delta_{0,1}(\tilde{H})\,,\quad {\cal A}(1\alpha b\shuffle c)\subset {\cal C}_{n-2}^{1}\Delta_{1,1}(\tilde{H})\,,\\
 &{\cal A}(1\alpha b\shuffle (c d))\subset {\cal C}_{n-2}^{1}\Delta_{1,2}(\tilde{H})\,,\dots 
\end{aligned}
\ee
where ${\cal C}^{1}$ denotes the first word of the list fixes as $1$, $\alpha$ is the permutation list, $b$, $c$ and $d$ are the words of the list.

The shuffle form $(p,q)$ maps to the shuffle tree ${\cal T}_{p,q}$, and the gravity amplitudes will be expressed as follows.
\be\label{eq-hopf}
{\cal M}_{n}=\sum _{i\in \{1,\ldots,n-3\}}  {\cal T}_{i_{1},\ldots,i_{k}} {\cal C}_{n}^{1}\Delta_{i_{1},\ldots,i_{k}}(\tilde{H}_{(n-1)n}\times \tilde{H}_{n(n-1)})\,,
\ee
where $i=i_{1}+\ldots i_{k}$, $\tilde{H}_{ab}$ denotes the Amplitudes space of the ${\cal A}(1\dots ab)$, ${\cal T}_{i_{1},\ldots,i_{k}}$ are shuffle weighted trees, and $\Delta(\tilde{H}_{(n-1)n}\times \tilde{H}_{n(n-1)})=\Delta(\tilde{H}_{(n-1)n})\cdot \Delta( \tilde{H}_{n(n-1)})$ since the product and the coproduct are compatible.

\section{The Collinear And Soft Limit}
\noindent At the tree level, the scattering amplitudes have the analytical structure. The pole behaviors come from the physical limits: soft and collinear limits. The gravity amplitudes have the universal soft factor \cite{Weinberg:1965nx}
 and universal splitting amplitudes \cite{Bern:1998sv}. These results can be re-derived from the tree formula \eqref{eq-KLT-tree-formula2} directly.

The soft limit is the momentum $p_{j}$ comes to zero, the tree graphs of the vertex $j$ have one degree, i.e., one edge with the other vertex contributes the soft factor to the tree formula.

The collinear limit is the $s_{ij}$ comes to zero. The tree graphs of the vertex $i$ connect with vertex $j$ contribute the splitting factor to the tree formula.

$i$,$j$ collinear:
	{\hbox{\lower 5.pt\hbox{\begin{tikzpicture}
    \draw[dotted,thick](-5pt,0)--(5pt,0)(30pt,0)--(40pt,0);
    \draw[Apricot](10pt,0)circle[radius=5pt]node{i}(25pt,0)circle[radius=5pt]node{j};
    \draw (15pt,0)--(20pt,0);
\end{tikzpicture}}}},\quad
$i$ soft:
	{\hbox{\lower 5.pt\hbox{\begin{tikzpicture}
    \draw[dotted,thick](-5pt,0)--(5pt,0);
    \draw[SkyBlue](10pt,0)circle[radius=5pt]node{i};
\end{tikzpicture}}}}.

\begin{figure}[htbp]
    \centering
    \subfigure[]{
    \begin{tikzpicture}[sibling distance =30pt,scale=0.45]
			\node[box] {1}
			child {node[box] {2}}{
				child {node[box,draw=SkyBlue] {\textcolor{SkyBlue}{3}}}};
				\end{tikzpicture} , \begin{tikzpicture}[sibling distance =25pt,scale=0.6]
			\node[box] {1}
			child {node[box] {2}}
			child {node[box,draw=SkyBlue] {\textcolor{SkyBlue}{3}}};
		\end{tikzpicture} 
    }
    \qquad \qquad
     \subfigure[]{
    	\begin{tikzpicture}[sibling distance =30pt,scale=0.45]
			\node[box] {1}
			child {node[box,draw=Apricot] {\textcolor{Apricot}{2}}}{
				child {node[box,draw=Apricot] {\textcolor{Apricot}{3}}}};
				\end{tikzpicture} , \begin{tikzpicture}[sibling distance =30pt,scale=0.45]
			\node[box] {1}
			child {node[box,draw=Apricot] {\textcolor{Apricot}{3}}}{
				child {node[box,draw=Apricot] {\textcolor{Apricot}{2}}}};
		\end{tikzpicture}
 }
    \caption{The tree graphs contribute the soft and collinear limit in the five-point. The (a) is in the soft limit, and (b) is in the collinear limit. }
    \label{fig-soft-collinear}
\end{figure}
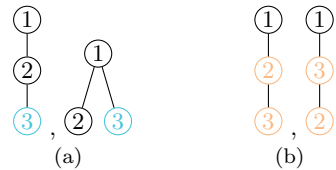

For example, in the five-point gravity amplitudes, the tree graphs contribute the soft factor in the soft limit ($p_{3}\to0$) are (a) in  Fig \ref{fig-soft-collinear}. Then the soft factor ${\cal S}^{gravity}$ is
\be
s_{23}\text{Soft}(2,3,4)\text{Soft}(2,3,5)+s_{13}\text{Soft}(1,3,4)\text{Soft}(1,3,5)\,,
\ee
where $\text{Soft}(a,j,b)$ is the soft factor in the gauge theory.

The tree graphs contribute the collinear factor in the collinear limit ($s_{23}\to0$) are (b) in  Fig \ref{fig-soft-collinear}. The collinear factor $\text{Split}^{gravity}$ is
\be
s_{23}\text{Split}(2,3)\text{Split}(2,3)\,,
\ee
where $\text{Split}(i,j)$ is the collinear factor in the gauge theory.\\
\\

\section{Conclusion}
\noindent We study the KLT relation in two aspects: the global and local aspects. The global KLT relation itself can emerge from the Permutohedron with the shuffle tree structure \eqref{eq-permutohedron}, which can be formed as the Hopf algebra \eqref{eq-hopf} while the elements of the KLT matrix have the Lie structure and binary tree representation \eqref{eq-binary} in the local aspect.  The geometric and algebraic structure of the KLT relation or KLT matrix deserves more attention and should have an equal status to the inverse of the KLT matrix. The direct study of the KLT relation will help us to discuss the double copy of the scattering amplitudes or some physical limits, such as the soft and collinear limits. We expect this work  will inspire more scholarship and reconsiderations of the KLT matrix from diverse perspectives.

\begin{acknowledgments}
\noindent We thank Hao Chen, Junding Chen, Bo Feng, Yibo Fu, Tingfei Li, Sebastan Mizera, Yaobo Zhang, Huaxing Zhu for useful comments and discussions. 
\end{acknowledgments}

\appendix
\section{\texorpdfstring{}{Appendix A: } The six-point KLT Permutohedron }\label{app}
\noindent In this appendix, we show the  six-point KLT Permutohedron in details. The six-point gravity amplitude can be derived from the tree formula~\eqref{eq-KLT-tree-formula}.
\begin{align}
	\mathcal{M}_6&={\hbox{\lower 30.pt\hbox{\begin{tikzpicture}[sibling distance =35pt,scale=0.4]
					\node[box] {1}
					child {node[box] {2}}{
						child {node[box] {3}}{
							child {node[box] {4}}}};
	\end{tikzpicture}}}}{\cal A}(1 2 3 4 5 6){\cal A}(1 2 3 4 6 5)+\text{perm}(234)\\
	&+{\hbox{\lower 30.pt\hbox{\begin{tikzpicture}[sibling distance =35pt,scale=0.5]
					\node[box] {1}
					child {node[box] {2}}{
						child {node[box] {3}}
						child {node[box] {4}}};
	\end{tikzpicture}}}}{\cal A}(1 2 (3\shuffle 4) 5 6){\cal A}(1 2 (3\shuffle 4) 6 5)\\&+{\hbox{\lower 30.pt\hbox{\begin{tikzpicture}[sibling distance =35pt,scale=0.5]
					\node[box] {1}
					child {node[box] {3}}{
						child {node[box] {2}}
						child {node[box] {4}}};
	\end{tikzpicture}}}}{\cal A}(1 3 (2\shuffle 4) 5 6){\cal A}(1 3 (2\shuffle 4) 6 5)\\
	&+{\hbox{\lower 30.pt\hbox{\begin{tikzpicture}[sibling distance =35pt,scale=0.5]
					\node[box] {1}
					child {node[box] {4}}{
						child {node[box] {2}}
						child {node[box] {3}}};
	\end{tikzpicture}}}}{\cal A}(1 4 (2\shuffle 3) 5 6){\cal A}(1 4 (2\shuffle 3) 6 5)\\
	&+{\hbox{\lower 30.pt\hbox{\begin{tikzpicture}[sibling distance =35pt,scale=0.5]
					\node[box] {1}
					child {node[box] {2}}
					child {node[box] {3}}{
						child {node[box] {4}}};
	\end{tikzpicture}}}}{\cal A}(1 (2\shuffle (3 4)) 5 6){\cal A}(1 (2\shuffle (3 4)) 6 5)+\text{perm}(34)\\
	&+{\hbox{\lower 30.pt\hbox{\begin{tikzpicture}[sibling distance =35pt,scale=0.5]
					\node[box] {1}
					child {node[box] {3}}
					child {node[box] {2}}{
						child {node[box] {4}}};
	\end{tikzpicture}}}}{\cal A}(1 (3\shuffle (2 4)) 5 6){\cal A}(1 (3\shuffle (2 4)) 6 5)+\text{perm}(24)\\
	&+{\hbox{\lower 30.pt\hbox{\begin{tikzpicture}[sibling distance =35pt,scale=0.5]
					\node[box] {1}
					child {node[box] {4}}
					child {node[box] {2}}{
						child {node[box] {3}}};
	\end{tikzpicture}}}}{\cal A}(1 (4\shuffle (2 3)) 5 6){\cal A}(1 (4\shuffle (2 3)) 6 5)+\text{perm}(23)\\
	&+{\hbox{\lower 10.pt\hbox{\begin{tikzpicture}[sibling distance =35pt,scale=0.5]
					\node[box] {1}
					child {node[box] {2}}
					child {node[box] {3}}
					child {node[box] {4}};
	\end{tikzpicture}}}}{\cal A}(1 (2\shuffle 3\shuffle 4)) 5 6){\cal A}(1 (2\shuffle 3\shuffle 4)) 6 5)\,,
\end{align}\\
where the perm is the permutation and $\shuffle$ is the shuffle operation. 

In the view of the KLT Permutohedron, $\mathcal{M}_6$ can be mapped from the $\mathcal{P}^{6}|_{3}$, which is restricted to the dimension $3$ part between permutation $\{1\alpha56\}$ and $\{1\alpha65\}$. All tree graphs come from the dimension $1$, $2$, and $3$ of the Permutohedron, made up of the KLT matrix in six points. The Permutohedron $\mathcal{P}^{6}|_{3}$ is a hexagonal prism. Each vertex represents a gauge amplitude.

\begin{minipage}{0.4\linewidth}	
	{\hbox{\lower 50.pt\hbox{\begin{tikzpicture}
					\draw (0:1)node(R1-0){}--(300:0.5)node(R1-315){}--(210:0.87)node(R1-225){}--(180:1)node(R1-180){};
					\draw[densely dashed,thick] (0:1)--(30:0.87)node(R1-45){}--(120:0.5)node(R1-135){}--(180:1);
					\begin{scope}[yshift=2.cm]
						\draw (0:1)node(R2-0){}--(300:0.5)node(R2-315){}--(210:0.87)node(R2-225){}--(180:1)node(R2-180){};
						\draw (0:1)--(30:0.87)node(R2-45){}--(120:0.5)node(R2-135){}--(180:1);
					\end{scope}
					
					\fill[orange!20] (R1-0.center)--(R1-45.center)--(R2-45.center)--(R2-0.center)--cycle;
					\fill[yellow!20] (R1-135.center)--(R1-180.center)--(R2-180.center)--(R2-135.center)--cycle;
					\fill[red!20] (R1-225.center)rectangle(R2-315.center);
					\draw[Black] (R2-180.center)--(R2-225.center);
					\draw[Black] (R2-315.center)--(R2-0.center);
					\draw[RedViolet,thick] (R1-0.center)--(R2-0.center);
					\draw[blue,thick] (R1-180.center)--(R2-180.center);
					\draw[LimeGreen,thick] (R1-225.center)--(R2-225.center);
					\draw[Green,thick] (R1-315.center)--(R2-315.center);
					\draw[Magenta,dashed,thick] (R1-45.center)--(R2-45.center);
					\draw[SkyBlue,dashed,thick] (R1-135.center)--(R2-135.center);
					\fill[blue] (R1-0.center)node[anchor=west]{$A$}circle[radius=2pt];
					\fill[violet] (R1-45.center)node[anchor=south west]{$B$}circle[radius=2pt];
					\fill[Emerald] (R1-135.center)node[anchor=south]{$C$}circle[radius=2pt];
					\fill[WildStrawberry] (R1-180.center)node[anchor=east]{$D$}circle[radius=2pt];
					\fill[YellowOrange] (R1-225.center)node[anchor=north]{$E$}circle[radius=2pt];
					\fill[RawSienna] (R1-315.center)node[anchor=west]{$F$}circle[radius=2pt];
					\fill[blue] (R2-0.center)node[anchor=west]{$A^\prime$}circle[radius=2pt];
					\fill[violet] (R2-45.center)node[anchor=south west]{$B^\prime$}circle[radius=2pt];
					\fill[Emerald] (R2-135.center)node[anchor=south]{$C^\prime$}circle[radius=2pt];
					\fill[WildStrawberry] (R2-180.center)node[anchor=east]{$D^\prime$}circle[radius=2pt];
					\fill[YellowOrange] (R2-225.center)node[anchor=north]{$E^\prime$}circle[radius=2pt];
					\fill[RawSienna] (R2-315.center)node[anchor=west]{$F^\prime$}circle[radius=2pt];
					\draw[dashed] (30:0.87)--(120:0.5);
					\draw[dashed] (180:1)--(120:0.5);
					\draw (R2-225.center)--(R2-315.center);
					\draw (R2-0.center)--(R2-45.center);
					\draw (R2-135.center)--(R2-180.center);
	\end{tikzpicture}}}}
\end{minipage}\hfill
\begin{minipage}{0.6\linewidth}
	A : ${\cal A}$(142356)\quad$A^\prime : {\cal A}$(142365)\\
	B : ${\cal A}$(143256)\quad$B^\prime : {\cal A}$(143265)\\
	C : ${\cal A}$(134256)\quad$C^\prime : {\cal A}$(134265)\\
	D : ${\cal A}$(132456)\quad$D^\prime : {\cal A}$(132465)\\
	E : ${\cal A}$(123456)\quad$E^\prime : {\cal A}$(123456)\\
	F : ${\cal A}$(124356)\quad$F^\prime : {\cal A}$(124356)		
\end{minipage}\\\vskip 1em
\begin{minipage}{0.4\linewidth}
	\begin{tikzpicture}
		\draw (0:1)node[right]{A}--(60:1)node[above]{B}--(120:1)node[above]{C}--(180:1)node[left]{D}--(240:1)node[below]{E}--(300:1)node[below]{F}--cycle;
		\draw[red] (0:1)--(240:1);
		\draw[green] (60:1)--(180:1);
		\draw[magenta] (120:1)--(240:1);
		\draw[violet] (60:1)--(300:1);
		\draw[orange] (180:1)--(300:1);
		\draw[teal] (0:1)--(120:1);
	\end{tikzpicture}
\end{minipage}\hfill
\begin{minipage}{0.6\linewidth}
	The left is  one facet of the Permutohedron. The edges between the top/down facets represent the one path graph. The rectangles represent one ${\cal T}_{(1,1)}$. The triangular prisms represent one ${\cal T}_{(1,2)}$.
\end{minipage}\\
The whole ${\cal P}^6|_{3}$ represents ${\cal T}_{(1,1,1)}$ \put(0,-10){ \begin{tikzpicture}[sibling distance =60pt,scale=0.4]
		\node[box] {1}
		child {node[box] {2}}
		child {node[box] {3}}
		child {node[box] {4}};
\end{tikzpicture}}\\
\begin{figure}[htbp]
	\label{fig-2}
	\centering  
	\includegraphics[scale=0.5]{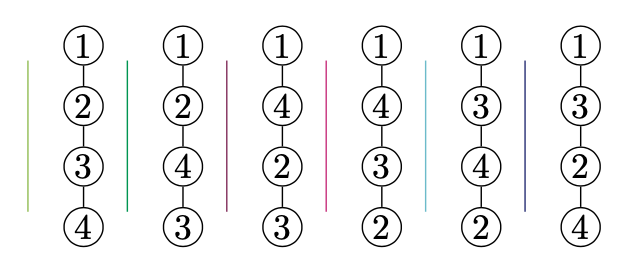}
	\caption{The edges between the top/down facets represent the one path graph.}
\end{figure}
\begin{figure}[htbp]
		\label{fig-3}
	\centering  
	\includegraphics[scale=0.5]{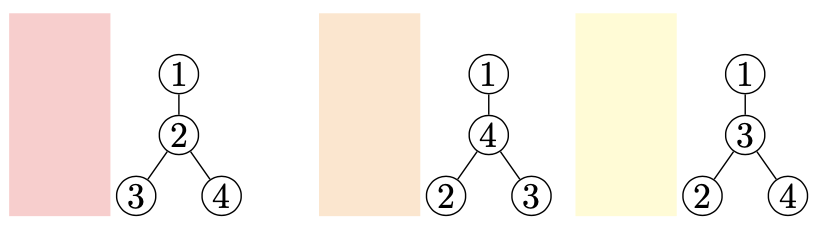}
	\caption{The rectangles  represent one ${\cal T}_{(1,1)}$.}
\end{figure}

\begin{figure}[htbp]
		\label{fig-4}
	\centering  
	\includegraphics[scale=0.5]{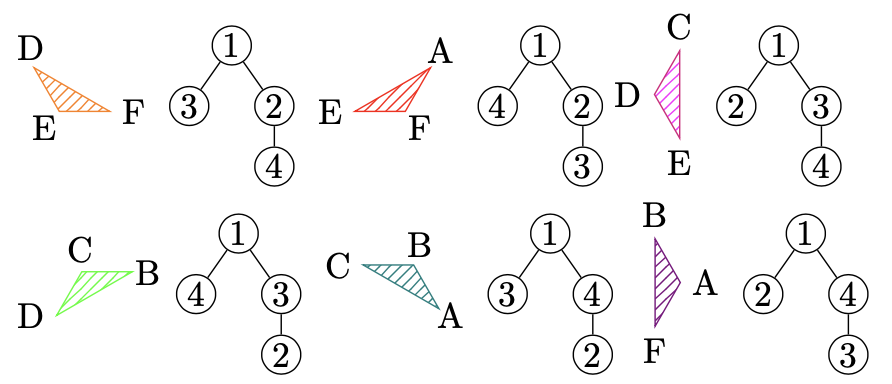}
	\caption{The triangular prisms represent one ${\cal T}_{(1,2)}$.}
\end{figure}

\bibliography{KLT}
\bibliographystyle{apsrev4-1}	

\end{document}